\documentclass[aps,prl,twocolumn,superscriptaddress,showpacs,showkeys,amsmath,amssymb,floatfix]{revtex4}
\usepackage{booktabs,graphicx,mathrsfs,verbatim}
\usepackage[usenames,dvipsnames]{color}
\usepackage{ulem}
\usepackage{color}
\usepackage[sort&compress]{natbib} 
\newcommand{\valencia}{\affiliation{Instituto de F{\'i}sica Corpuscular, CSIC-Universidad de Valencia, E-46071 Valencia, Spain}}
\newcommand{\osakadep}{\affiliation{Department of Physics, Osaka University, Toyonaka, Osaka 560-0043, Japan}}
\newcommand{\bordeaux}{\affiliation{Centre d'Etudes Nucl{\'e}aires de Bordeaux Gradignan, CNRS/IN2P3 - Universit{\'e} Bordeaux 1, 33175 Gradignan Cedex, France}}
\newcommand{\surrey}{\affiliation{Department of Physics, University of Surrey, Guildford GU2 7XH, Surrey, UK}}
\newcommand{\debrecen}{\affiliation{Inst. of Nuclear Research of the Hung. Acad. of Sciences, Debrecen, H-4026, Hungary}}
\newcommand{\instanbul}{\affiliation{Department of Physics, Istanbul University, Istanbul, 34134, Turkey}}
\newcommand{\caen}{\affiliation{Grand Acc{\'e}l{\'e}rateur National d'Ions Lourds, BP 55027, F-14076 Caen, France}}
\newcommand{\osakarcnp}{\affiliation{Research Center for Nuclear Physics, Osaka University, Ibaraki, Osaka 567-0047, Japan}}
\newcommand{\chile}{\affiliation{Comisi{\'o}n Chilena de Energ{\'i}a Nuclear, Casilla 188-D, Santiago, Chile}}
\newcommand{\belgium}{\affiliation{SCK.CEN, Boeretang 200, 2400 Mol, Belgium}}
\newcommand{\argonne}{\affiliation{Physics Division, Argonne National Laboratory, Argonne, Illinois 60439, USA}}

\begin{document} 

\title{Observation of the $\beta$-delayed $\gamma$-proton decay of $^{56}$Zn and its impact on the Gamow Teller strength evaluation}

\author{S.~E.~A.~Orrigo}\email{sonja.orrigo@ific.uv.es}\valencia
\author{B.~Rubio}\valencia
\author{Y.~Fujita}\osakadep\osakarcnp
\author{B.~Blank}\bordeaux
\author{W.~Gelletly}\surrey
\author{J.~Agramunt}\valencia
\author{A.~Algora}\valencia\debrecen
\author{P.~Ascher}\bordeaux
\author{B.~Bilgier}\instanbul
\author{L.~C{\'a}ceres}\caen
\author{R.~B.~Cakirli}\instanbul
\author{H.~Fujita}\osakarcnp
\author{E.~Ganio{\u{g}}lu}\instanbul
\author{M.~Gerbaux}\bordeaux
\author{J.~Giovinazzo}\bordeaux
\author{S.~Gr{\'e}vy}\bordeaux
\author{O.~Kamalou}\caen
\author{H.~C.~Kozer}\instanbul
\author{L.~Kucuk}\instanbul
\author{T.~Kurtukian-Nieto}\bordeaux
\author{F.~Molina}\valencia\chile
\author{L.~Popescu}\belgium
\author{A.~M.~Rogers}\argonne
\author{G.~Susoy}\instanbul
\author{C.~Stodel}\caen
\author{T.~Suzuki}\osakarcnp
\author{A.~Tamii}\osakarcnp
\author{J.~C.~Thomas}\caen


\begin{abstract}
We report the observation of a very exotic decay mode at the proton drip-line, the $\beta$-delayed $\gamma$-proton decay, clearly seen in the $\beta$ decay of the $T_z$ = -2 nucleus $^{56}$Zn. Three $\gamma$-proton sequences have been observed after the $\beta$ decay. Here this decay mode, already observed in the $sd$-shell, is seen for the first time in the $fp$-shell. Both $\gamma$ and proton decays have been taken into account in the estimation of the Fermi (F) and Gamow Teller (GT) strengths. Evidence for fragmentation of the Fermi strength due to strong isospin mixing is found. 
\end{abstract}



\maketitle


The study of the properties of nuclei far from stability lies at one of the main frontiers of modern nuclear physics. The appearance of new radioactive decay modes is among the marked changes as the nuclei become more and more weakly bound \cite{Blank2008,Pfutzner2012}. Instead of the familiar $\alpha$ and $\beta$ radioactivity or $\gamma$ de-excitation we find close to the proton drip-line proton ($p$-) \cite{Hofmann1982,Klepper1982} and $2p$-radioactivity \cite{Giovinazzo2002,Pfutzner2002}. As the states involved become particle-unbound, we observe $\beta$-delayed proton decay, $\beta$-delayed neutron decay and $\beta$-delayed fission \cite{Kuznetsov1966,Kuznetsov1967,Andreyev2013} in heavy systems, and many more decay channels open in lighter nuclei, see for instance $^{11}$Li \cite{Madurga2008}. In medium-heavy nuclei, such as $^{56}$Zn, fewer decay channels are expected but close to the drip-lines there may be open channels not yet observed.

Among many possible observables for nuclear structure, the $\beta$-decay strengths provide important testing grounds for nuclear structure theories far from stability. The mechanism of $\beta$ decay is well understood and dominated by allowed Fermi (F) and Gamow-Teller (GT) transitions. A successful description of the nuclear structure of the states involved should provide good predictions for the corresponding transition strengths $B$(F) and $B$(GT). The F transition feeds the Isobaric Analogue State (IAS), which forms an isospin multiplet with the ground state (gs) of the parent nucleus.

In this Letter we present a decay mode that has been observed only rarely before and can strongly affect the conventional way to study $\beta$ decay close to the drip-line.

The $B$(F) and $B$(GT) values are estimated from the $\beta$ feeding, the half-life $T_{1/2}$ and decay energy for each branch. For medium to heavy nuclei, close to stability, the de-excitation proceeds via $\beta$-delayed $\gamma$ decay and the $\beta$ feeding is obtained from the balance between the $\gamma$ intensity feeding and de-exciting each level. As the nuclei become more exotic, the particle separation energies decrease and the strong interaction causes decay to occur via $\beta$-delayed particle emission. In proton-rich nuclei proton decay is expected to dominate for states well above ($>$1 MeV) the proton separation energy $S_{p}$ and the $\beta$ feeding is readily inferred from the intensities of the proton peaks. However, the de-excitation of the IAS via proton decay is usually isospin forbidden. In this case competition between $\beta$-delayed proton emission and $\beta$-delayed $\gamma$ de-excitation becomes possible even at energies well above $S_{p}$ \cite{Dossat2007,Bhattacharya2008,Fujita2013}. Thus one has to take into account the intensities of both the proton and $\gamma$ emission to estimate $B$(F). Normally this only affects $B$(F) but, in special circumstances as in the present case, $B$(GT) is also affected. 

We report the results of a study of the $T_{z} = -2 \rightarrow -1, \beta^{+}$ decay of $^{56}$Zn to $^{56}$Cu [$T_{z} = (N-Z)/2$ is the third component of the isospin quantum number $T$], where we observe competition between $\beta$-delayed proton and $\gamma$ emission in states well above $S_{p}$. Moreover we observe $\beta$-delayed $\gamma$ rays that populate proton-unbound levels that subsequently decay by proton emission. This observation is very important because it does affect the conventional way to determine $B$(GT) near the proton drip-line, where the general opinion until now was that $B$(GT) is simply deduced from the intensity of the proton peaks. Here instead, to determine $B$(GT) properly, the intensity of the proton transitions has to be corrected for the amount of indirect feeding coming from the $\gamma$ de-excitation. Although similar cases were suggested in the $sd$-shell \cite{Wrede2009,Pfutzner2012} and observed in the decay of $^{32}$Ar \cite{Bhattacharya2008}, how this new decay mode affects the determination of $B$(GT) has never been discussed.


Prior to the present work little was known about the decay of $^{56}$Zn and the excited states of its daughter $^{56}$Cu. $\beta$-delayed protons were observed \cite{Dossat2007}, but not $\beta$-delayed $\gamma$ rays. The present experiment was motivated by a comparison with the mirror charge exchange (CE) reaction on $^{56}$Fe \cite{HFujita2010}. Indeed $\beta$ decay and CE studies are complementary and, assuming the isospin symmetry, they can be combined to determine the absolute $B$(GT) values up to high excitation energies \cite{Taddeucci1987,Fujita2005,Fujita2011}. Hence as precise a determination of $B$(GT) as possible is important.


The $\beta$-decay experiment was performed at the LISE3 facility of GANIL \cite{Mueller1991} in 2010, using a $^{58}$Ni$^{26+}$ primary beam with an average intensity of 3.7 e$\mu$A. This beam, accelerated to 74.5 MeV/nucleon, was fragmented on a 200 $\mu$m thick natural Ni target. The fragments were selected by the LISE3 separator and implanted into a Double-Sided Silicon Strip Detector (DSSSD), surrounded by four EXOGAM Ge clovers for $\gamma$ detection. The DSSSD was 300 $\mu$m thick and had 16 X and 16 Y strips with a pitch of 3 mm, defining 256 pixels. They were used to detect both the implanted fragments and subsequent charged-particle ($\beta$s and protons) decays. For this purpose, two parallel electronic chains were used having different gains. An implantation event was defined by simultaneous signals in both a silicon $\Delta E$ detector located upstream and the DSSSD. The implanted ions were identified by combining the energy loss signal in the $\Delta E$ detector and the Time-of-Flight (ToF) defined as the time difference between the cyclotron radio-frequency and the $\Delta E$ signal. Decay events were defined as giving a signal above threshold (typically 50-90 keV) in the DSSSD and no coincident signal in the $\Delta E$ detector.

\begin{figure}[!t]
  \centering
  \includegraphics[height=1.\columnwidth, angle=90]{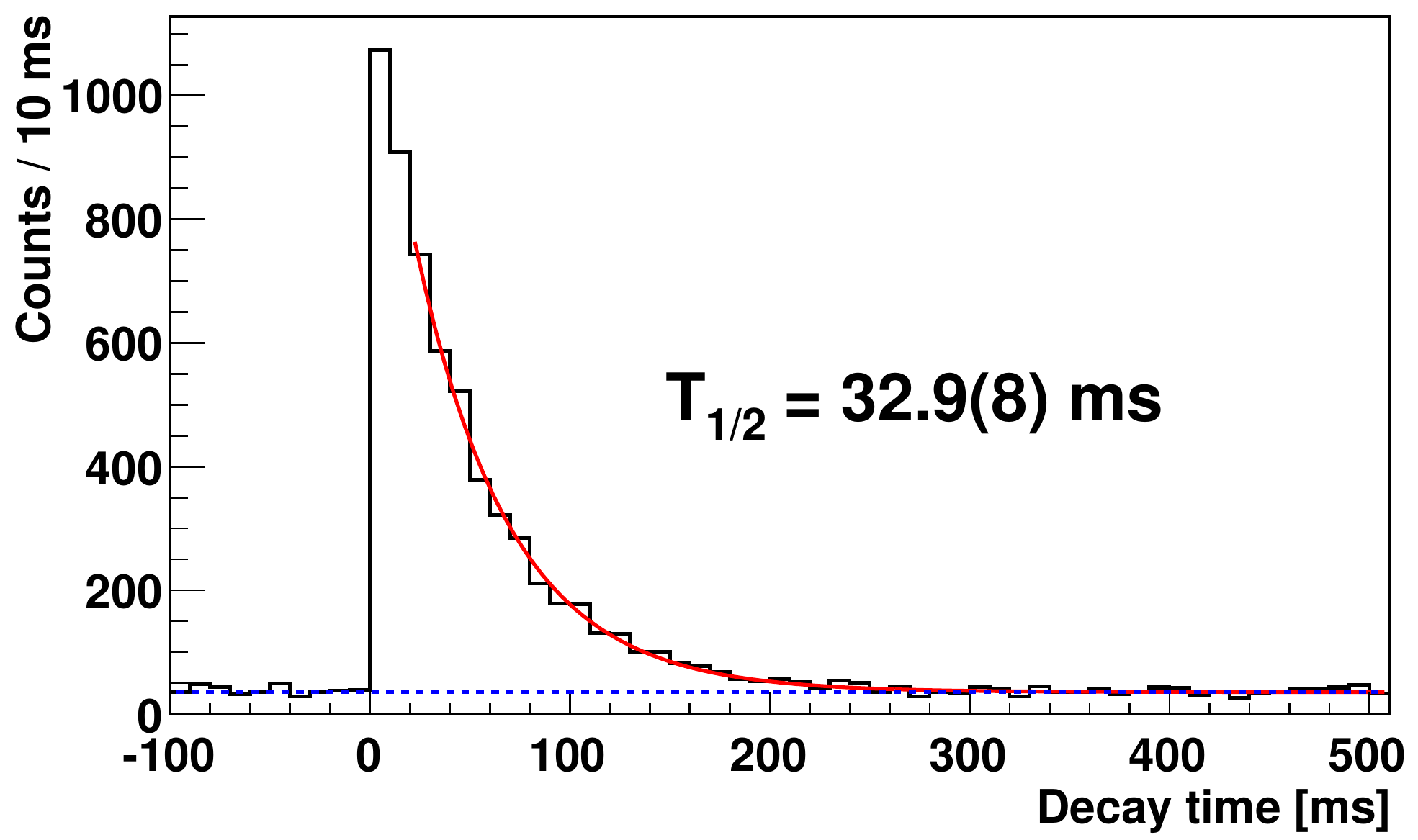}
  \vspace{-5.0 mm}
	\caption{Spectrum of the time correlations between each proton decay (DSSSD energy $>$ 800 keV) and all the $^{56}$Zn implants.}
	\label{halflife}
  \vspace{-5.0 mm}
\end{figure}

The $^{56}$Zn ions were selected by setting gates off-line on the $\Delta E$-ToF matrix. The total number of implanted $^{56}$Zn nuclei was 8.9x10$^{3}$. The $\it{correlation~time}$ is defined as the time difference between a decay event in a given pixel of the DSSSD and any implantation signal that occurred before and after it in the same pixel that satisfied the conditions required to identify the nuclear species. The proton decays were selected by setting an energy threshold above 800 keV (removing the $\beta$ decays, see the discussion about the DSSSD spectrum) and looking for correlated $^{56}$Zn implants. This procedure ensured that all the true correlations were taken into account. However many random correlations were also included producing, as expected, a large constant background. Fig. \ref{halflife} shows the correlation-time spectrum for $^{56}$Zn. The data were fitted with a function including the $\beta$ decay of $^{56}$Zn and a constant background. A half-life ($T_{1/2}$) of 32.9(8) ms was obtained for $^{56}$Zn, in agreement with \cite{Dossat2007}.

\begin{figure}[!t]
	\begin{minipage}{1.0\linewidth}
		\centering
		\includegraphics[height=1.\columnwidth, angle=90]{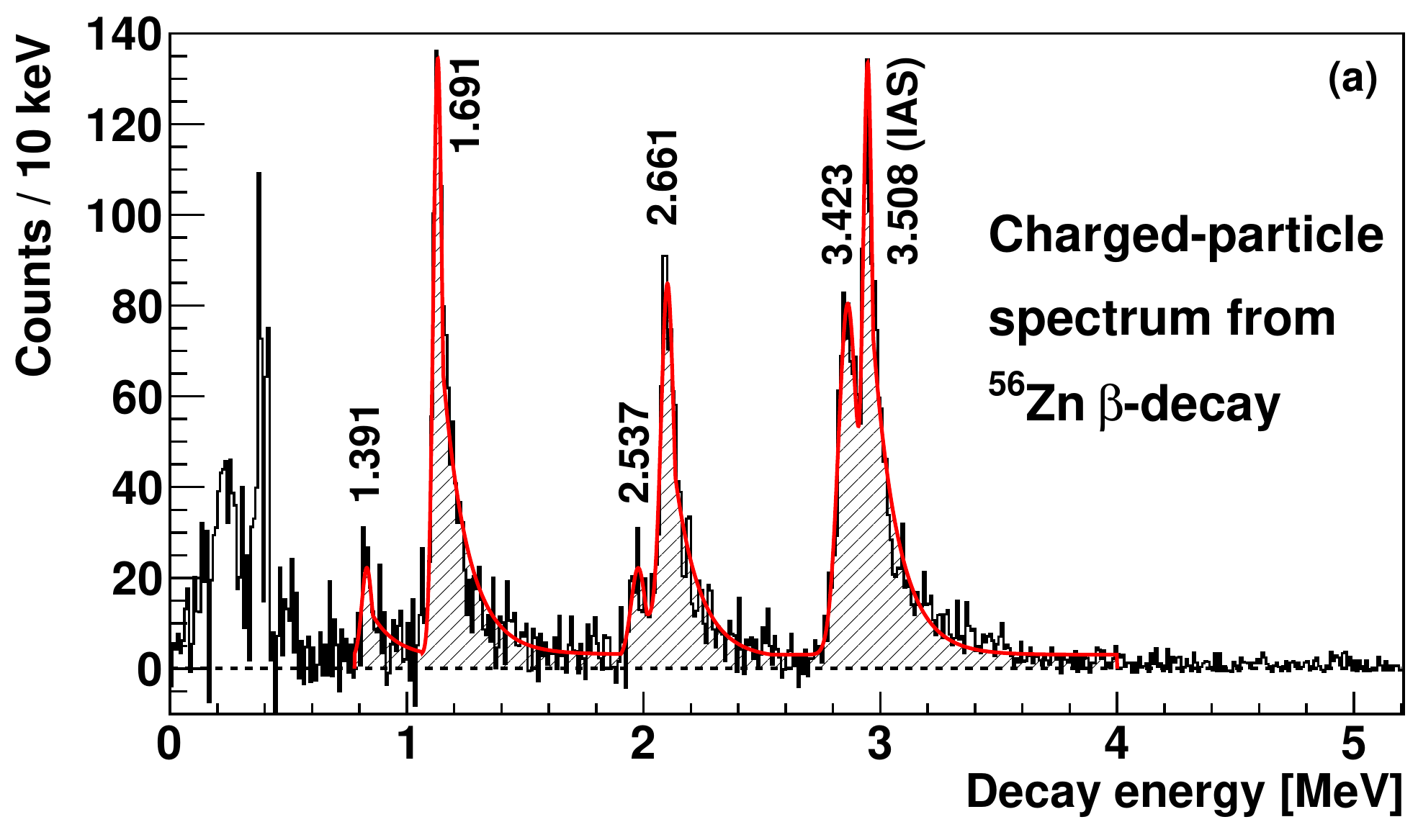}
	\end{minipage}
	\begin{minipage}{1.0\linewidth}
		\centering
    \includegraphics[height=1.\columnwidth, angle=90]{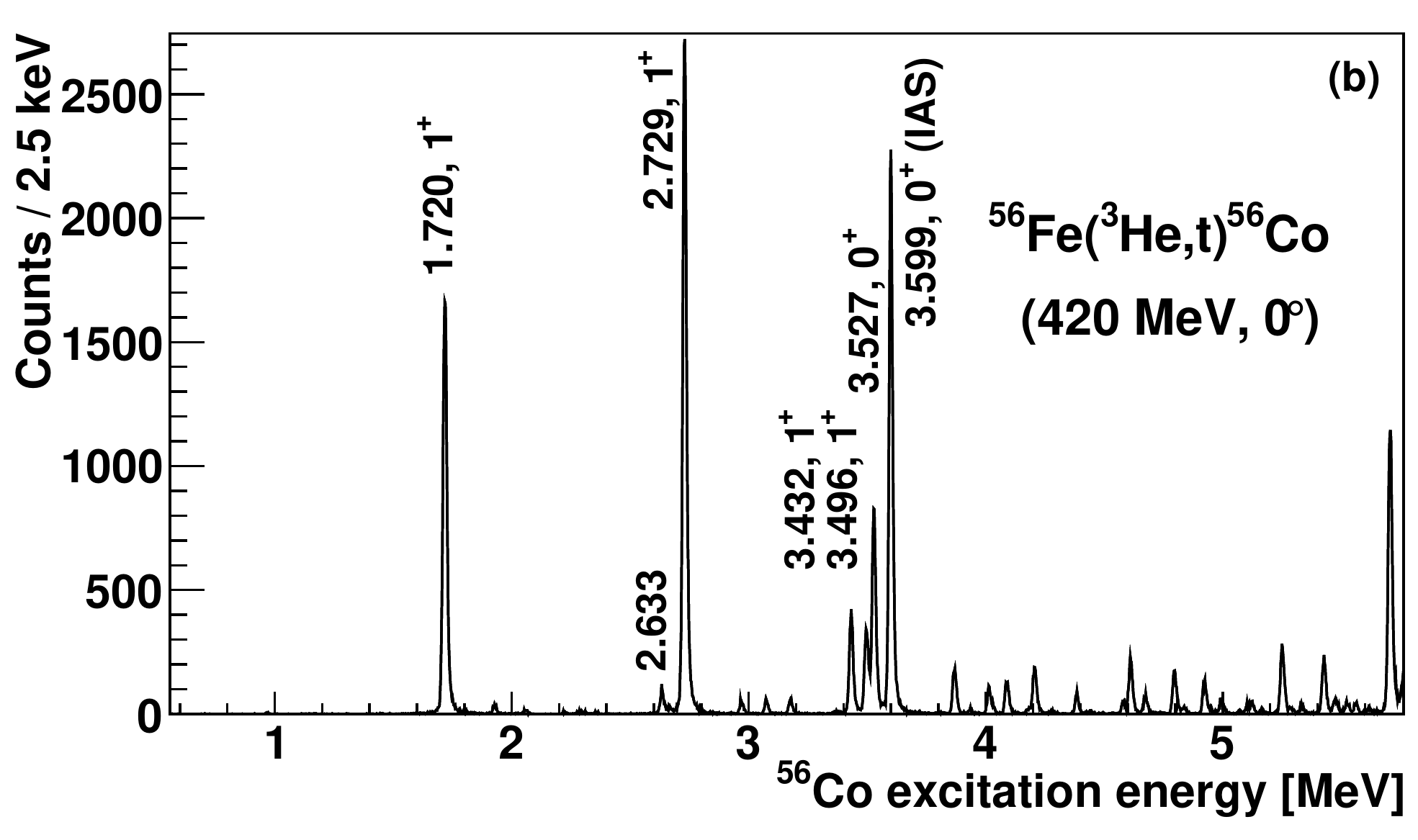}
	  \vspace{-5.0 mm}
		\caption{(a) Charged-particle spectrum measured in the DSSSD for decay events correlated with $^{56}$Zn implants. The peaks are labeled according to the corresponding excitation energies in $^{56}$Cu. (b) $^{56}$Fe($^{3}$He,$t$)$^{56}$Co reaction spectrum \cite{HFujita2010}. Peaks are labeled by the excitation energies in $^{56}$Co.}
		\label{p+CE-spectra}
	\end{minipage}
	\vspace{-5.0 mm}
\end{figure}

The charged-particle spectrum measured in the DSSSD for decays associated with $^{56}$Zn implants is shown in Fig. \ref{p+CE-spectra}(a). It was formed as in \cite{Dossat2007} and calibrated using an $\alpha$-particle source and the $^{53}$Ni peaks of known energy.

Two kinds of state are expected to be populated in the $\beta$ decay of $^{56}$Zn to $^{56}$Cu: the $T$ = 2, $J^{\pi}$= 0$^{+}$ IAS, and a number of $T$ = 1, 1$^{+}$ states. From the comparison with the mirror nucleus $^{56}$Co, all of these states will lie above $S_p^\#$ = 560(140) keV \cite{Audi2003} ($\#$ means from systematics), thus they will decay by proton emission. Indeed most of the strength in Fig. \ref{p+CE-spectra}(a) is interpreted as $\beta$-delayed proton emission to the $^{55}$Ni$_{gs}$. We attribute the broad bump below 800 keV to $\beta$ particles that are not in coincidence with protons. The proton peaks seen above 800 keV are labeled in terms of the excitation energies in $^{56}$Cu. The large uncertainty of $\pm$140 keV in the $^{56}$Cu level energies comes from the uncertainty in the estimated $S_p^\#$. The proton decay of the IAS is identified as the peak at 3508 keV, as in \cite{Dossat2007}. The energy resolution for protons is 70 keV FWHM. The achievable resolution is limited by the summing with the coincident $\beta$ particles, which also affects the lineshape of the peak. MonteCarlo simulations confirm that the lineshape is well approximated by a Gaussian plus an exponential high-energy tail (as in \cite{Dossat2007}). This shape was used for the fit shown in Fig. \ref{p+CE-spectra}(a).

Figure \ref{p+CE-spectra}(b) shows the peaks in the triton spectrum corresponding to the excitation of states in $^{56}$Co, populated in the mirror $T_{z} = +2 \rightarrow +1$, $\beta^{-}$-type CE reaction on the stable $T_{z}$ = +2 target $^{56}$Fe. The spectrum was obtained with a high resolution of $\sim30$ keV in the $^{56}$Fe($^{3}$He,$t$)$^{56}$Co reaction at RCNP Osaka \cite{HFujita2010}. Figs. \ref{p+CE-spectra}(a) and \ref{p+CE-spectra}(b) have been aligned in excitation energy in $^{56}$Cu and $^{56}$Co. There is a good correspondence between states in the mirror nuclei $^{56}$Cu and $^{56}$Co. The 3508 keV IAS and the states at 2661 and 1691 keV in $^{56}$Cu correspond to the 3599 keV IAS and the levels at 2729 and 1720 keV in $^{56}$Co, respectively. 

In Fig. \ref{p+CE-spectra}(a) one can see that the 3423 keV peak is broader than others. We suggest that this peak corresponds to the three states seen in $^{56}$Co in Fig. \ref{p+CE-spectra}(b). One of the $^{56}$Co states is the $T$ = 1, $0^{+}$ level at 3527 keV. It mixes with the IAS and takes 28\% of the Fermi strength \cite{Dzubay1970,HFujita2010} in the ($^{3}$He,$t$) reaction. 

The peaks at 2537 and 1391 keV probably correspond to the 2633 and 1451 keV levels in $^{56}$Co, respectively. The former is only weakly populated in the CE reaction, while the latter is the 0$^{+}$ anti-analogue state \cite{Kampp1978} and is not observed in CE. Consequently the 2537 keV and 1391 keV states are expected to have little feeding in the $\beta$ decay and the observed protons indicate that they are indirectly populated by $\gamma$ decay from the levels above.

\begin{figure}[!b]
	\centering
  \vspace{-5.0 mm}
	\includegraphics[height=1.\columnwidth, angle=90]{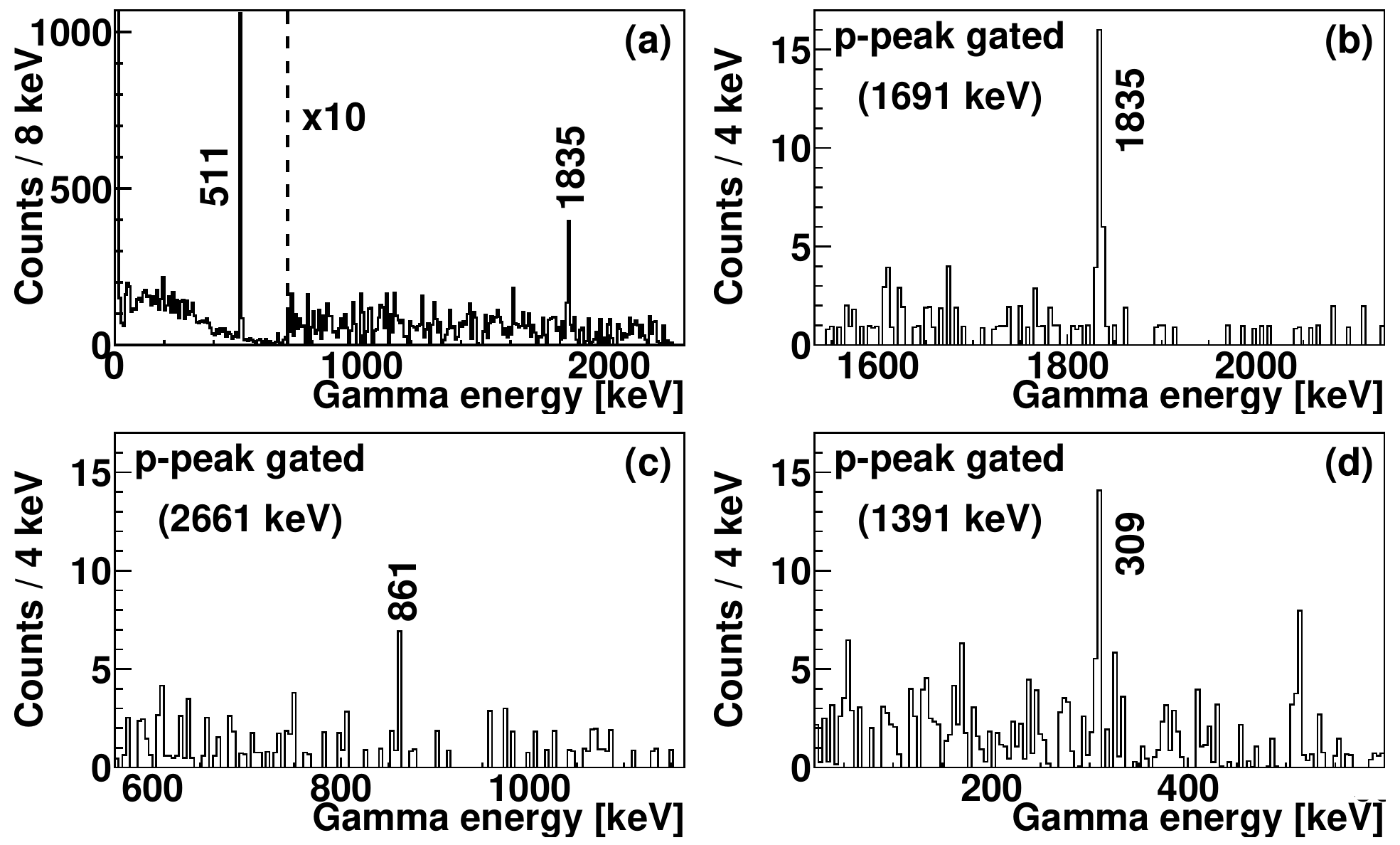}
	\caption{$\gamma$-ray spectra in coincidence with (a) all charged particles and protons from the levels at (b) 1691, (c) 2661 and (d) 1391 keV.}
	\label{g-spectrum}
	\vspace{-5.0 mm}
\end{figure}

The proton decay of the 3508 keV, $T=2$ IAS in $^{56}$Cu to the $T=1/2$, $^{55}$Ni$_{gs}$ is normally expected to be isospin forbidden \cite{Dossat2007,Fujita2013}, which makes the competing $\gamma$ de-excitation possible. Fig. \ref{g-spectrum}(a) shows the $\gamma$-ray spectrum measured in coincidence with the decays correlated with the $^{56}$Zn implants, after removal of the background of random correlations as in the DSSSD spectrum. A $\gamma$ line was observed at 1834.5(10) keV, in agreement with the energy difference between the 3508 and 1691 keV $^{56}$Cu states [1817(15) keV], therefore this $\gamma$ line is attributed to the transition connecting these levels. Further confirmation arises from the 1835 keV line being in coincidence with the proton decay from the 1691 keV level [Fig. \ref{g-spectrum}(b)]. Moreover, the half-life associated with the 1835 keV $\gamma$ ray, $T_{1/2}$ = 27(8) ms, is in good agreement with the $^{56}$Zn half-life.

Other cases of $\gamma$ decay from an IAS above $S_p$ have been observed in this mass region \cite{Dossat2007,Fujita2013}. The particular circumstance here is that the 1691 keV level is also proton-unbound, consequently the rare and exotic $\beta$-delayed $\gamma$-proton decay has been observed for the first time in the $fp$-shell. Similar cases were investigated in the $sd$-shell \cite{Bhattacharya2008,Wrede2009,Pfutzner2012}. Our observation cannot be interpreted as a $\beta$-delayed proton-decay to an excited state of $^{55}$Ni, followed by a $\gamma$ decay to the $^{55}$Ni$_{gs}$, because the state populated would have an energy of 1835 keV, while the first excited state in $^{55}$Ni lies at 2089 keV. 

Imposing coincidence conditions on the various proton peaks in Fig. \ref{p+CE-spectra}(a), two additional $\gamma$ rays are observed. The first one, seen at 861 keV [Fig. \ref{g-spectrum}(c)], corresponds to the de-excitation from the 3508 keV IAS to the 2661 keV state. The second lies at 309 keV [Fig. \ref{g-spectrum}(d)] and is interpreted as the transition connecting the 1691 and 1391 keV states.

\begin{figure}[!t]
	\centering
	\includegraphics[height=1.\columnwidth, angle=90]{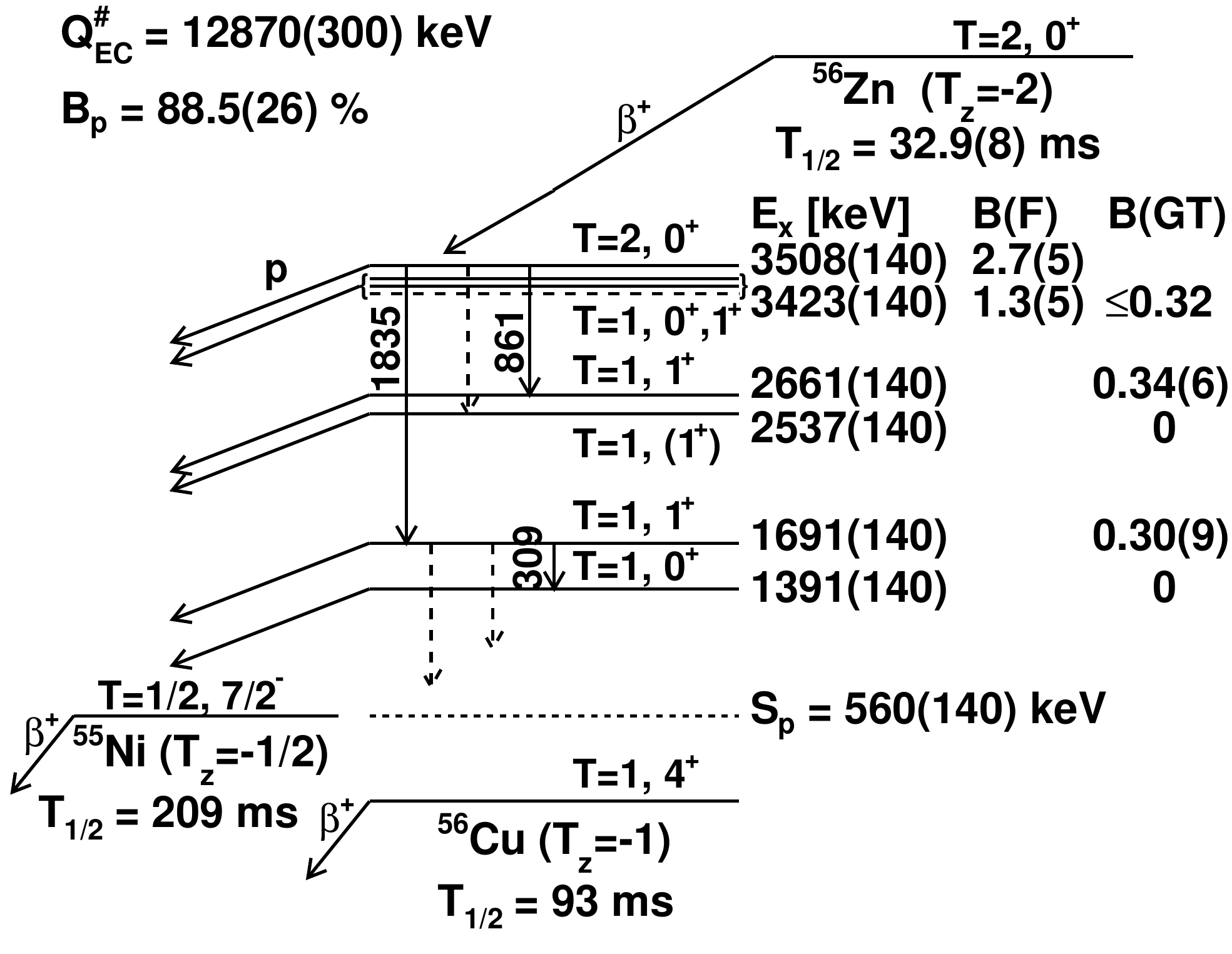}
  \caption{$^{56}$Zn decay scheme deduced from the present experiment. Observed proton or $\gamma$ decays are indicated by solid lines. Transitions corresponding to those seen in the mirror $^{56}$Co nucleus are shown by dashed lines. The error of 140 keV comes from the uncertainty in $S_p^\#$.}
	\label{decay-scheme}
	\vspace{-5.0 mm}
\end{figure}

All of our observations are summarized in the $^{56}$Zn decay scheme in Fig. \ref{decay-scheme}. Solid lines indicate experimentally observed proton and $\gamma$ decays; dashed lines represent transitions seen in the mirror $^{56}$Co nucleus. Three cases of $\beta$-delayed $\gamma$-proton emission have been established experimentally, involving the $\gamma$ rays at 1835, 861 and 309 keV.


Table \ref{Table1} shows the energies of the proton and $\gamma$ peaks, and their intensities deduced from the areas of the peaks. For a proper determination of $B$(F) and $B$(GT), the $\beta$ feeding to each $^{56}$Cu level was estimated from the proton and $\gamma$ intensities, taking into account the amount of indirect feeding produced by the $\gamma$ de-excitation. In doing that we have used the intensities of the observed $\gamma$ lines and estimates based on the $\gamma$ de-excitation pattern in the mirror $^{56}$Co nucleus \cite{Kampp1978}. Assuming 100\% DSSSD efficiency for both implants and protons (see Fig. 5 in \cite{Dossat2007}), a total proton branching ratio $B_p$ = 88.5(26)\% is obtained by comparing the total number of $^{56}$Zn implants with the number of observed protons above 800 keV [Fig. \ref{p+CE-spectra}(a)]. A reasonable systematic error of 20 $\mu$m in the implantation depth would lead to a proton efficiency still very close to 100\%. This systematic error is included in the quoted uncertainty. The missing 11.5(26)\% is attributed to the $\beta$-delayed $\gamma$ emission from the 1691 keV level (in analogy with $^{56}$Co), where the estimated partial proton half-life is $t_{1/2}\sim~10^{-14}$ s, an order-of-magnitude where the $\gamma$ de-excitation can compete with the proton emission. It is estimated that the $\gamma$ decays represent 56(6)\% and 66(22)\% of the total decays from the 3508 keV IAS and 1691 keV state, respectively. 


The measured $T_{1/2}$, $\beta$ feedings $I_\beta$ and $B_p$ were used to determine the $B$(F) and $B$(GT) values. For completeness the results obtained taking $Q_{EC}^\#$ and $S_p^\#$ from \cite{Audi2003} and from \cite{Audi2012} are shown in Table \ref{Table2}. The differences in the derived values of $B$(F) and $B$(GT) are very small. The energies of the corresponding levels in $^{56}$Cu and $^{56}$Co differ by less than 100 keV when using \cite{Audi2003}, and by $\sim400$ keV when using \cite{Audi2012}, therefore we use \cite{Audi2003} in the discussion.

The total Fermi transition strength has to be $|N-Z|$ = 4. The $^{56}$Cu IAS at 3508 keV has a Fermi strength $B$(F) = 2.7(5). The missing strength, 1.3(5), has to be hidden in the broad peak at 3423 keV. This is a confirmation that the $^{56}$Cu IAS is fragmented and thus part of the feeding of the 3423 keV level (assuming it contains two or three unresolved levels) corresponds to the Fermi transition and the remaining part of it to the GT transition. The isospin impurity $\alpha^2$ (defined as in \cite{HFujita2010}) and the off-diagonal matrix element of the charge-dependent part of the Hamiltonian $\left\langle{H_{c}}\right\rangle$, responsible for the isospin mixing of the 3508 keV IAS (0$^{+}$, $T=2$) and the 0$^{+}$ part of the 3423 keV level ($T=1$), are computed according to two-level mixing. For $^{56}$Cu it was found that $\left\langle{H_{c}}\right\rangle$ = 40(23) keV and $\alpha^2$ = 33(10)\%, similar to the values obtained in $^{56}$Co \cite{HFujita2010}.

An interesting and puzzling open question is, considering that the isospin mixing is quite large in  $^{56}$Cu, why we are still observing the $\gamma$ de-excitation from the IAS in competition with the (in principle) much faster and now partially-allowed proton decay ($t_{1/2}\sim~10^{-18}$ s).

\begin{table}[!t]
	\caption{Proton energies, $\gamma$-ray energies, and their intensities (normalized to 100 decays) for the decay of $^{56}$Zn. *IAS.}
	\label{Table1}
	\centering
	\begin{ruledtabular}
	  \begin{tabular}{c c c c}
		  $E_p$(keV) & $I_p$(\%) & $E_{\gamma}$(keV) & $I_{\gamma}$(\%)\\ \hline
		  2948(10)* & 18.8(10) & 1834.5(10) & 16.3(49)\\
		  2863(10) & 21.2(10) & 861.2(10) & 2.9(10)\\
		  2101(10) & 17.1(9) & 309.0(10) & - \\
		  1977(10) & 4.6(8) & & \\
		  1131(10) & 23.8(11) & & \\
		  831(10) & 3.0(4) & & \\
	  \end{tabular}
	\end{ruledtabular}
  \vspace{-5.0 mm}
\end{table}

\begin{table}[!t]
	\caption{$\beta$ feedings, Fermi and Gamow Teller transition strengths to the $^{56}$Cu levels in the $\beta^{+}$ decay of $^{56}$Zn calculated using $^a$\cite{Audi2003} and $^b$\cite{Audi2012}. *IAS.}
	\label{Table2}
	\centering
	\begin{ruledtabular}
	  \begin{tabular}{c c c c c c c}
		  $I_\beta$(\%) & $E^a$(keV) & $B$(F)$^a$ & $B$(GT)$^a$ & $E^b$(keV) & $B$(F)$^b$ & $B$(GT)$^b$\\ \hline
		  43(5) & 3508(140)* & 2.7(5) & & 3138(200)* & 2.5(5) &\\
		  21(1) & 3423(140) & 1.3(5) & $\leq$0.32 & 3053(200) & 1.2(5) & $\leq$0.31\\
		  14(1) & 2661(140) & & 0.34(6) & 2291(200) & & 0.31(6)\\
		  0 & 2537(140) & & 0 & 2167(200) & & 0\\
		  22(6) & 1691(140) & & 0.30(9) & 1321(200) & & 0.28(8)\\
		  0 & 1391(140) & & 0 & 1021(200) & & 0\\
	  \end{tabular}
	\end{ruledtabular}
  \vspace{-5.0 mm}
\end{table}

The $B$(F) and $B$(GT) values obtained for the $^{56}$Cu levels (Table \ref{Table2}) are in agreement with the corresponding values in $^{56}$Co \cite{HFujita2010}. For the GT strength calculations it was assumed that the 1391 and 2537 keV levels get no direct feeding, thus the corresponding $B$(GT) = 0. This assumption, compatible with our data, is based on the comparison with the CE data [Fig. \ref{p+CE-spectra}(b)], where the two mirror levels at 1451 and 2633 keV are only very weakly populated if at all. The 3423 keV level surely has $B$(GT) $>$ 0 because the related broad proton peak contains at least one 1$^{+}$ state.


In summary, the $^{56}$Zn half-life and decay scheme, with absolute $B$(F) and $B$(GT) strengths, have been established. Evidence for fragmentation of the IAS in $^{56}$Cu due to isospin mixing is seen. Competition between $\beta$-delayed proton and $\gamma$ emission is observed in two states in $^{56}$Cu above $S_p$. Moreover, in three cases, we have also observed $\beta$-delayed $\gamma$-proton emission for the first time in the $fp$-shell. This exotic decay mode will strongly affect the conventional determination of $B$(GT) in heavier, more exotic systems with $T_{z} \leq -3/2$ (where studies are planned at present and future radioactive beam facilities), in which it may become a common decay mode. This is an important message because it shows that the decay may not be entirely dominated by particle emission and the use of $\gamma$ detectors is necessary.


\begin{acknowledgments}
This work was supported by the Spanish MICINN grants FPA2008-06419-C02-01, FPA2011-24553; CPAN Consolider-Ingenio 2010 Programme CSD2007-00042; MEXT, Japan 18540270 and 22540310; Japan-Spain coll. program of JSPS and CSIC; Istanbul University Scientific Research Projects, Num. 5808; UK Science and Technology Facilities Council (STFC) Grant No. ST/F012012/1; Region of Aquitaine. R.B.C. acknowledges support by the Alexander von Humboldt foundation and the Max-Planck-Partner Group. We acknowledge the EXOGAM collaboration for the use of their clover detectors.
\end{acknowledgments}




\begin{thebibliography}{50}

\bibitem{Blank2008}
B. Blank and M. J. G. Borge, Prog. Part. Nucl. Phys. \textbf{60}, 403 (2008).

\bibitem{Pfutzner2012}
M. Pf\"utzner, M. Karny, L. V. Grigorenko and K. Riisager, Rev. Mod. Phys.\textbf{84}, 567 (2012).

\bibitem{Hofmann1982}
S. Hofmann, W. Reisdorf, G. M{\"{u}}nzenberg, F. Hessberger, J. Schneider, P. Armbruster, Z. Phys. A \textbf{305}, 111 (1982).

\bibitem{Klepper1982}
O. Klepper et al., Z. Phys. A \textbf{305}, 125 (1982).

\bibitem{Giovinazzo2002} 
J. Giovinazzo et al., Phys. Rev. Lett. \textbf{89}, 102501 (2002).

\bibitem{Pfutzner2002} 
M. Pf{\"{u}}tzner et al., Eur. Phys. J. A \textbf{14}, 279 (2002).

\bibitem{Kuznetsov1966}
V. I. Kuznetsov, N. K. Skobelev and G. N. Flerov, Yad. Fiz. \textbf{4}, 279 (1966) [Sov. J. Nucl. Phys. \textbf{4}, 202 (1967)].

\bibitem{Kuznetsov1967}
V. I. Kuznetsov, N. K. Skobelev and G. N. Flerov, Yad. Fiz. \textbf{5}, 271 (1967) [Sov. J. Nucl. Phys. \textbf{5}, 191 (1967)].

\bibitem{Andreyev2013}
A. N. Andreyev et al., Phys. Rev. C \textbf{87}, 014317 (2013). 

\bibitem{Madurga2008}
M. Madurga et al., Nucl. Phys. A \textbf{810}, 1 (2008).

\bibitem{Dossat2007}
C. Dossat et al., Nucl. Phys. A \textbf{792}, 18 (2007).

\bibitem{Bhattacharya2008}
M. Bhattacharya et al., Phys. Rev. C \textbf{77}, 065503 (2008).

\bibitem{Fujita2013}
Y. Fujita et al., Phys. Rev. C \textbf{88}, 014308 (2013).

\bibitem{Wrede2009}
C. Wrede et al., Phys. Rev. C \textbf{79}, 045808 (2009).

\bibitem{HFujita2010}
H. Fujita et al., Phys. Rev. C \textbf{88}, 054329 (2013).

\bibitem{Taddeucci1987} 
T. N. Taddeucci et al., Nucl. Phys. A \textbf{469}, 125 (1987).

\bibitem{Fujita2005} 
Y. Fujita et al., Phys. Rev. Lett. \textbf{95}, 212501 (2005).

\bibitem{Fujita2011} 
Y. Fujita, B. Rubio and W. Gelletly, Prog. Part. Nucl. Phys. \textbf{66}, 549 (2011).

\bibitem{Mueller1991}
A. C. Mueller and R. Anne, Nucl. Instr. and Meth. B \textbf{56-57}, 559 (1991).

\bibitem{Audi2003}
G. Audi et al., Nucl. Phys. A \textbf{729}, 1 (2003).

\bibitem{Dzubay1970}
T.D. Dzubay et al., Nucl. Phys. A \textbf{142}, 488 (1970).

\bibitem{Kampp1978}
W.D. Kampp et al., Z. Phys. A \textbf{288}, 167 (1978).

\bibitem{Audi2012}
G. Audi et al., Chin. Phys. C \textbf{36}, 1157 (2012).

\end{thebibliography}
\end{document}